\documentclass[aps,prl,twocolumn,showpacs,superscriptaddress,groupedaddress]{revtex4}  
\usepackage{graphicx}  
\usepackage{dcolumn}   
\usepackage{bm}        
\usepackage{amssymb}   
\usepackage{subfigure}
\usepackage{array}
\def\met{\mbox{${\hbox{$E$\kern-0.6em\lower-.1ex\hbox{/}}}_T$}} 

\begin{document}

\hspace{5.2in} \mbox{Fermilab-Pub-13-361-E}


\title{Measurement of the muon charge asymmetry in
{\bm{$p\overline{p} \rightarrow W + X \rightarrow \mu\nu + X$}}
events at {\bm{$\sqrt{s} = 1.96$}}~TeV}

\affiliation{LAFEX, Centro Brasileiro de Pesquisas F\'{i}sicas, Rio de Janeiro, Brazil}
\affiliation{Universidade do Estado do Rio de Janeiro, Rio de Janeiro, Brazil}
\affiliation{Universidade Federal do ABC, Santo Andr\'e, Brazil}
\affiliation{University of Science and Technology of China, Hefei, People's Republic of China}
\affiliation{Universidad de los Andes, Bogot\'a, Colombia}
\affiliation{Charles University, Faculty of Mathematics and Physics, Center for Particle Physics, Prague, Czech Republic}
\affiliation{Czech Technical University in Prague, Prague, Czech Republic}
\affiliation{Institute of Physics, Academy of Sciences of the Czech Republic, Prague, Czech Republic}
\affiliation{Universidad San Francisco de Quito, Quito, Ecuador}
\affiliation{LPC, Universit\'e Blaise Pascal, CNRS/IN2P3, Clermont, France}
\affiliation{LPSC, Universit\'e Joseph Fourier Grenoble 1, CNRS/IN2P3, Institut National Polytechnique de Grenoble, Grenoble, France}
\affiliation{CPPM, Aix-Marseille Universit\'e, CNRS/IN2P3, Marseille, France}
\affiliation{LAL, Universit\'e Paris-Sud, CNRS/IN2P3, Orsay, France}
\affiliation{LPNHE, Universit\'es Paris VI and VII, CNRS/IN2P3, Paris, France}
\affiliation{CEA, Irfu, SPP, Saclay, France}
\affiliation{IPHC, Universit\'e de Strasbourg, CNRS/IN2P3, Strasbourg, France}
\affiliation{IPNL, Universit\'e Lyon 1, CNRS/IN2P3, Villeurbanne, France and Universit\'e de Lyon, Lyon, France}
\affiliation{III. Physikalisches Institut A, RWTH Aachen University, Aachen, Germany}
\affiliation{Physikalisches Institut, Universit\"at Freiburg, Freiburg, Germany}
\affiliation{II. Physikalisches Institut, Georg-August-Universit\"at G\"ottingen, G\"ottingen, Germany}
\affiliation{Institut f\"ur Physik, Universit\"at Mainz, Mainz, Germany}
\affiliation{Ludwig-Maximilians-Universit\"at M\"unchen, M\"unchen, Germany}
\affiliation{Panjab University, Chandigarh, India}
\affiliation{Delhi University, Delhi, India}
\affiliation{Tata Institute of Fundamental Research, Mumbai, India}
\affiliation{University College Dublin, Dublin, Ireland}
\affiliation{Korea Detector Laboratory, Korea University, Seoul, Korea}
\affiliation{CINVESTAV, Mexico City, Mexico}
\affiliation{Nikhef, Science Park, Amsterdam, the Netherlands}
\affiliation{Radboud University Nijmegen, Nijmegen, the Netherlands}
\affiliation{Joint Institute for Nuclear Research, Dubna, Russia}
\affiliation{Institute for Theoretical and Experimental Physics, Moscow, Russia}
\affiliation{Moscow State University, Moscow, Russia}
\affiliation{Institute for High Energy Physics, Protvino, Russia}
\affiliation{Petersburg Nuclear Physics Institute, St. Petersburg, Russia}
\affiliation{Instituci\'{o} Catalana de Recerca i Estudis Avan\c{c}ats (ICREA) and Institut de F\'{i}sica d'Altes Energies (IFAE), Barcelona, Spain}
\affiliation{Uppsala University, Uppsala, Sweden}
\affiliation{Lancaster University, Lancaster LA1 4YB, United Kingdom}
\affiliation{Imperial College London, London SW7 2AZ, United Kingdom}
\affiliation{The University of Manchester, Manchester M13 9PL, United Kingdom}
\affiliation{University of Arizona, Tucson, Arizona 85721, USA}
\affiliation{University of California Riverside, Riverside, California 92521, USA}
\affiliation{Florida State University, Tallahassee, Florida 32306, USA}
\affiliation{Fermi National Accelerator Laboratory, Batavia, Illinois 60510, USA}
\affiliation{University of Illinois at Chicago, Chicago, Illinois 60607, USA}
\affiliation{Northern Illinois University, DeKalb, Illinois 60115, USA}
\affiliation{Northwestern University, Evanston, Illinois 60208, USA}
\affiliation{Indiana University, Bloomington, Indiana 47405, USA}
\affiliation{Purdue University Calumet, Hammond, Indiana 46323, USA}
\affiliation{University of Notre Dame, Notre Dame, Indiana 46556, USA}
\affiliation{Iowa State University, Ames, Iowa 50011, USA}
\affiliation{University of Kansas, Lawrence, Kansas 66045, USA}
\affiliation{Louisiana Tech University, Ruston, Louisiana 71272, USA}
\affiliation{Northeastern University, Boston, Massachusetts 02115, USA}
\affiliation{University of Michigan, Ann Arbor, Michigan 48109, USA}
\affiliation{Michigan State University, East Lansing, Michigan 48824, USA}
\affiliation{University of Mississippi, University, Mississippi 38677, USA}
\affiliation{University of Nebraska, Lincoln, Nebraska 68588, USA}
\affiliation{Rutgers University, Piscataway, New Jersey 08855, USA}
\affiliation{Princeton University, Princeton, New Jersey 08544, USA}
\affiliation{State University of New York, Buffalo, New York 14260, USA}
\affiliation{University of Rochester, Rochester, New York 14627, USA}
\affiliation{State University of New York, Stony Brook, New York 11794, USA}
\affiliation{Brookhaven National Laboratory, Upton, New York 11973, USA}
\affiliation{Langston University, Langston, Oklahoma 73050, USA}
\affiliation{University of Oklahoma, Norman, Oklahoma 73019, USA}
\affiliation{Oklahoma State University, Stillwater, Oklahoma 74078, USA}
\affiliation{Brown University, Providence, Rhode Island 02912, USA}
\affiliation{University of Texas, Arlington, Texas 76019, USA}
\affiliation{Southern Methodist University, Dallas, Texas 75275, USA}
\affiliation{Rice University, Houston, Texas 77005, USA}
\affiliation{University of Virginia, Charlottesville, Virginia 22904, USA}
\affiliation{University of Washington, Seattle, Washington 98195, USA}
\author{V.M.~Abazov} \affiliation{Joint Institute for Nuclear Research, Dubna, Russia}
\author{B.~Abbott} \affiliation{University of Oklahoma, Norman, Oklahoma 73019, USA}
\author{B.S.~Acharya} \affiliation{Tata Institute of Fundamental Research, Mumbai, India}
\author{M.~Adams} \affiliation{University of Illinois at Chicago, Chicago, Illinois 60607, USA}
\author{T.~Adams} \affiliation{Florida State University, Tallahassee, Florida 32306, USA}
\author{J.P.~Agnew} \affiliation{The University of Manchester, Manchester M13 9PL, United Kingdom}
\author{G.D.~Alexeev} \affiliation{Joint Institute for Nuclear Research, Dubna, Russia}
\author{G.~Alkhazov} \affiliation{Petersburg Nuclear Physics Institute, St. Petersburg, Russia}
\author{A.~Alton$^{a}$} \affiliation{University of Michigan, Ann Arbor, Michigan 48109, USA}
\author{A.~Askew} \affiliation{Florida State University, Tallahassee, Florida 32306, USA}
% ...
%
%
\collaboration{The D0 Collaboration\footnote{with visitors from
$^{a}$Augustana College, Sioux Falls, SD, USA,
$^{b}$The University of Liverpool, Liverpool, UK,
$^{c}$DESY, Hamburg, Germany,
$^{d}$Universidad Michoacana de San Nicolas de Hidalgo, Morelia, Mexico
$^{e}$SLAC, Menlo Park, CA, USA,
$^{f}$University College London, London, UK,
$^{g}$Centro de Investigacion en Computacion - IPN, Mexico City, Mexico,
$^{h}$Universidade Estadual Paulista, S\~ao Paulo, Brazil,
$^{i}$Karlsruher Institut f\"ur Technologie (KIT) - Steinbuch Centre for Computing (SCC)
and
$^{j}$Office of Science, U.S. Department of Energy, Washington, D.C. 20585, USA.
}} \noaffiliation
\vskip 0.25cm

\date{September 10, 2013}

\begin{abstract}
We present a measurement of the muon charge asymmetry from the decay of the $W$ boson via $W \rightarrow \mu\nu$ using 7.3~fb$^{-1}$ of integrated luminosity collected with the D0 detector at the Fermilab Tevatron Collider at $\sqrt{s} = 1.96$~TeV. The muon charge asymmetry is presented in two kinematic regions in muon transverse momentum and event missing transverse energy:  ($p^{\mu}_{T} > 25 {\text{~GeV, }} \met > 25 {\text{~GeV}}$) and ($p^{\mu}_{T} > 35 {\text{~GeV, }}  \met > 35 {\text{~GeV}}$). The measured asymmetries are compared with theory predictions made using three parton distribution function sets.  The predictions do not describe the data well for $p^{\mu}_{T} > 35$~GeV, $\met > 35$~GeV, and larger values of muon pseudorapidity.
\end{abstract}

\maketitle

A measurement of the muon charge asymmetry from the decays of $W^{\pm}$ bosons produced in $p\overline{p}$ collisions provides information that constrains the parton distribution functions (PDFs) of the $u$ and $d$ quarks in the proton. At the Fermilab Tevatron Collider, $W^{+}$ ($W^{-}$) bosons are primarily produced by interactions between valence $u$ ($d$) quarks in the proton and valence $\overline{d}$ ($\overline{u}$) antiquarks in the antiproton. On average, $u$ quarks carry more of the proton momentum than $d$ quarks~\cite{Berger}. Therefore, $W^{+}$ bosons tend to be produced with momenta along the direction of the proton, while $W^{-}$ bosons tend to be produced with momenta along the direction of the antiproton.  The $W$ boson asymmetry is defined as \begin{equation}
  A_{W}(y) = \frac{\frac{\mathrm{d}\sigma}{\mathrm{d}y}(W^{+}) - \frac{\mathrm{d}\sigma}{\mathrm{d}y}(W^{-})}{\frac{\mathrm{d}\sigma}{\mathrm{d}y}(W^{+}) + \frac{\mathrm{d}\sigma}{\mathrm{d}y}(W^{-})},
\end{equation}
 where ${\mathrm{d}\sigma}/{\mathrm{d}y}(W^{\pm})$ is the differential cross section for $p\overline{p}\rightarrow W^{\pm} + X$, and $y$ is the $W$ boson rapidity.  Assuming an SU(3) symmetric quark-antiquark sea, that the quark PDFs in the proton are equal to the antiquark PDFs in the antiproton, and that valence quark interactions are the dominant source of $W$ boson production, 
\begin{equation}
  A_{W}(y) \approx \frac{\frac{d(x_2)}{u(x_2)} - \frac{d(x_1)}{u(x_1)}}{\frac{d(x_2)}{u(x_2)} + \frac{d(x_1)}{u(x_1)}},
\end{equation}
where $u(x)$ and $d(x)$ are the PDFs for the up and down quarks, and $x_1$ and $x_2$ are the momentum fractions carried by the interacting quarks in the proton and the antiproton, respectively.  At leading order, the quark momentum fractions and the $W$ boson rapidity are related by 
\begin{equation}
x_{1(2)} = \frac{M_W}{\sqrt{s}} e^{+(-) y},
\end{equation}
where $M_W$ is the $W$ boson mass.

In the $W\rightarrow\mu\nu$ process, the muon charge asymmetry is a convolution of the $W$ boson production asymmetry with the asymmetry from the {\sl V--A} decay of the $W$ boson.  At higher lepton $p_T$, the {\sl V--A}  contribution is smaller, so that the muon charge asymmetry is larger and closer to the $W$ boson asymmetry; at higher muon rapidity, the {\sl V--A} contribution is larger, and the muon asymmetry is significantly smaller than the $W$ boson asymmetry.  Since the {\sl V--A} interaction is well understood, the muon charge asymmetry can be used to probe the $u$ and $d$ quark PDFs.

The lepton charge asymmetry in the decay of $W$ bosons produced in $p\overline{p}$ collisions has been measured by both the CDF~\cite{CDFLeptonAsym1995,CDFLeptonAsym1998,CDFLeptonAsym2005} and D0~\cite{Sinjini,David} Collaborations. The most recent lepton charge asymmetry measurement from the D0 Collaboration was done in the electron channel using 0.75~fb$^{-1}$ of integrated luminosity.  The CDF Collaboration performed a direct measurement of the $W$ boson production asymmetry using 1~fb$^{-1}$ of integrated luminosity~\cite{CDFWAsym2009}.  The lepton charge asymmetry in $pp$ collisions, where $W$ boson production involves antiquarks from the proton sea, was measured by the ATLAS~\cite{ATLASAsym} and CMS~\cite{CMSAsym} Collaborations at the LHC using integrated luminosities of 31~pb$^{-1}$ and 36~pb$^{-1}$, respectively.  Here, we present a measurement of the muon charge asymmetry using 7.3~fb$^{-1}$ of $p\overline{p}$ data at $\sqrt{s} = 1.96$~TeV. This measurement supersedes our previous result in the muon channel \cite{Sinjini} and provides constraints on the ratio of the $u$ and $d$ quark PDFs in the region $0.005 \lesssim x \lesssim 0.3$ at $Q^{2} \approx M^{2}_{W}$~\cite{Sinjini}, where $Q$ is the momentum transfer.

In this analysis, the muon charge asymmetry is measured as a function of muon pseudorapidity $\eta^\mu$ where $\eta = -\ln[\tan(\theta/2)]$, and $\theta$ is the polar angle with respect to the proton beam direction. In the massless limit, $\eta$ is equal to the rapidity. The muon charge asymmetry is defined as
\begin{equation}
  A_{\mu}(\eta^\mu) = \frac{\frac{\mathrm{d}\sigma}{\mathrm{d}\eta}(\mu^{+}) - \frac{\mathrm{d}\sigma}{\mathrm{d}\eta}(\mu^{-})}{\frac{\mathrm{d}\sigma}{\mathrm{d}\eta}(\mu^{+}) + \frac{\mathrm{d}\sigma}{\mathrm{d}\eta}(\mu^{-})},
\end{equation}
 where ${\mathrm{d}\sigma}/{\mathrm{d}\eta}(\mu^{\pm})$ is the differential cross section for $p\overline{p}\rightarrow W^{\pm}\rightarrow\mu^{\pm}\nu + X$.

The D0 detector consists of a central tracking system, a calorimeter, and a muon system. The central tracking system contains a silicon microstrip tracker (SMT) and a central fiber tracker (CFT) and is located within a 1.9~T superconducting solenoidal magnet. The maximum coverage in $|\eta_{\text{det}}|$ for the SMT is 3.0; it is 2.5 for the CFT, where $|\eta_{\text{det}}|$ is the pseudorapidity measured from the center of the detector. The liquid-argon and uranium calorimeter has a central section covering $|\eta_{\text{det}}| < 1.1$ and two end caps extending the coverage to $|\eta_{\text{det}}|$ $\approx 4.2$. The muon system consists primarily of three layers of scintillation trigger counters and tracking detectors: one layer before a 1.8~T magnetized iron toroid and two layers outside the magnet; coverage extends to $|\eta_{\text{det}}| \approx 2.0$. A detailed description of the D0 detector is given in Refs.~\cite{d0det,L0-nim}; muon reconstruction and identification are described in Ref.~\cite{muon-nim}. 

We use two data samples: the full Run~IIa (2002 -- 2006) data set with 1.0~fb$^{-1}$ of integrated luminosity and 6.3~fb$^{-1}$ of integrated luminosity \cite{lum-note} collected during Run~IIb (2006 -- 2010).  Both integrated luminosities are after application of  the relevant data quality requirements.  The two data samples are analyzed independently because of changes in the detector configuration and the increased instantaneous luminosity during Run~IIb.  Candidate events are selected using a set of single-muon triggers that require the muon transverse momentum $p^{\mu}_T$ to be at least 10~GeV. The widest $|\eta_{\text{det}}|$ coverage of the single-muon triggers for Run~IIa (Run~IIb) data is 2.0 (1.6). Events are selected offline by requiring the $p\bar{p}$ collision vertex to have at least two tracks and to be located within 60 cm of the center of the detector along the beam direction. Muon candidates are required to lie within the acceptance of the detector and to be spatially matched to a track in the central tracking system with $p^{\mu}_{T} > 25$~GeV. The distance along the beam direction between the matched muon track and the $p\overline{p}$ vertex must be less than 2~cm. Muons are required to be isolated from other energy depositions.  The total transverse momentum of the tracks in a cone of radius $\Delta R = \sqrt{(\Delta\eta)^{2} + (\Delta\phi)^{2}}= 0.5$ around the matched central track must be less than 2.5 GeV,  where $\phi$ is the azimuthal angle, and the $p_T$ of the central track is excluded.  The total transverse energy measured in the calorimeter in a hollow cone of inner radius 0.1 and outer radius 0.5 around the muon must be less than 2.5 GeV.  The muons must be separated from any jet \cite{jetref} with transverse energy $E^{\text{jet}}_{T} > 15$~GeV by a distance $\Delta R > 0.5$. 

In general, the longitudinal momenta of neutrinos cannot be measured at a hadron collider. The neutrino transverse energy is inferred from the missing transverse energy \met, which is the negative vector sum of the transverse energy deposited in the calorimeter and the muon transverse momentum. Selected events must have $\met > 25$~GeV and transverse mass $M_{T} > 50$~GeV, where $M_{T} = \sqrt{2p^{\mu}_{T}\met(1 - \cos\Delta\phi)}$, and $\Delta\phi$ is the azimuthal angle between the muon and the \met\ in the plane transverse to the beam. There are 2.8 million events satisfying all of the selection criteria.  

The asymmetry measurement is made as a function of $\eta^\mu$ for two inclusive kinematic regions: ($p^{\mu}_{T} > 25$~GeV, $\met > 25$~GeV) and ($p^{\mu}_{T} > 35$~GeV, $\met > 35$~GeV).  The use of the same selection requirements for $p^{\mu}_{T}$ and $\met$ reduces the dependence of the muon asymmetry on the $W$ boson $p_T$.  The asymmetry is calculated as
\begin{equation}
  A_\mu = \frac{(1+kg-g)N^{+} - (k-kg+g)N^{-}}
           {(1-kg-g)N^{+} + (k-kg-g)N^{-}},
\label{Eq:Asym2}
\end{equation}
where $g$ is the muon charge misidentification probability, $k = \varepsilon^{+}/\varepsilon^{-}$ is the relative efficiency for positive and negative muons, and $N^+$ ($N^-$) is the number of positive (negative) muon events corrected for backgrounds and integrated luminosity, as described below.  The Run~IIa and Run~IIb data samples have different acceptances and detector efficiencies, therefore, each ($p^{\mu}_{T}$,\met,$\eta^\mu$,Run) region is treated independently.  All average values given below are over both data samples in the ($p^{\mu}_{T} > 25$~GeV, $\met > 25$~GeV) kinematic region.

Misidentification of the muon charge dilutes the muon charge asymmetry. We measure the probability that the muon charge is determined incorrectly using a tag-and-probe method and $Z\rightarrow\mu\mu$ events. We require one muon, the tag, to satisfy the selection criteria used for the signal, while the second muon, the probe, must satisfy looser requirements.  The dimuon mass is required to be above 50~GeV.  The probe is then tested against the selection requirement in question, and the ratio of the number of passing probes to the number of total probes is the efficiency of the selection requirement. The charge misidentification probability is the ratio of the number of tag-probe events in which the two muons have the same charge to the total number of events.  Uncertainty due to background in the $Z\rightarrow\mu\mu$ sample is taken into account.  The average muon charge misidentification probability is $g = (0.06 \pm 0.01)$\% for $|\eta^\mu| < 2$.

In the D0 detector, the directions of the magnetic fields in the solenoidal and toroidal magnets are reversed regularly to reduce any asymmetry due to the detector.  However, the portions of data in each polarity combination are not identical.  Approximately 50.2\% (49.1\%) of the data was collected with one solenoid (toroid) polarity and 49.8\% (50.9\%) with the opposite polarity. 
Therefore, any residual charge asymmetry from the tracking system where $p^\mu_{T}$ is measured will affect the muon charge asymmetry measurement. To correct for any charge asymmetry due to the detector, we weight the data so that all four polarity combinations have the same integrated luminosity.  The systematic uncertainty due to the magnet polarity weighting is determined from the uncertainty on the luminosity measurement excluding the uncertainty on the total inelastic cross section.

In principle, the acceptances and efficiencies are independent of muon charge since the directions of the magnetic fields in the solenoidal magnet and the magnetized iron are reversed frequently.  However, although the overall $p^{\mu}_T$ distributions for positive and negative muons are identical for $W$ boson decay, the $p^{\mu}_{T}$ distributions for positive and negative muons are not identical for a given $\eta^\mu$ region, especially at high $|\eta^\mu|$.  Since the muon identification efficiency depends on $p^{\mu}_T$, a relative efficiency correction must be included.  The muon reconstruction efficiency, the tracking efficiency, the isolation efficiency, and the trigger efficiency as functions of $\eta^\mu$,  $p^{\mu}_{T}$, and instantaneous luminosity are found using the dimuon data set and the tag-and-probe method.  The isolation efficiency is also found as a function of $\Delta R$ between the muon and the nearest jet and as a function of the $\eta_{\text{det}}$ position of the muon within the CFT.  On average, the muon reconstruction efficiency is $(74 \pm 1)$\%. The average tracking efficiency is $(90 \pm 1)$\%. The average isolation efficiency is $(86 \pm 4)$\%, and the average trigger efficiency is $(66 \pm 1)$\%. The product of the four efficiencies defines the overall muon efficiency with an average of $(38 \pm 2)$\%.  The overall efficiency is used to determine $k$, which ranges from 1.00 for $0.0 < |\eta^\mu| < 0.2$ to 1.01 for $1.8 < |\eta^\mu| < 2.0$.

The main background in the analysis is from electroweak processes: $Z\rightarrow\mu\mu$ where one muon is not reconstructed and $W\rightarrow\tau\nu$ and $Z\rightarrow\tau\tau$ where a tau lepton decays to a muon. The electroweak background is estimated using Monte Carlo (MC) samples generated with {\sc pythia}~\cite{pythia}, processed with a detailed simulation of the D0 detector based on {\sc geant}~\cite{geant}, and reconstructed using the same reconstruction code as used for the data. The fractions of each background source in the $W\rightarrow\mu\nu$ candidate samples are ($5.5 \pm 0.4$)\% for $Z\rightarrow\mu\mu$, ($1.6 \pm 0.1$)\% for $W\rightarrow\tau\nu$, and ($0.09 \pm 0.01$)\% for $Z\rightarrow\tau\tau$ for ($p^{\mu}_{T} > 25$~GeV, $\met > 25$~GeV). 

\begin{figure}[!ht]
\includegraphics[scale=0.5]{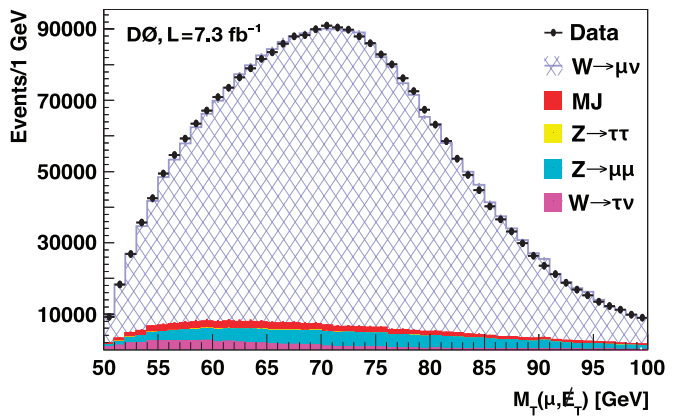}
\caption{[color online].  The transverse mass of selected events with $p^\mu_T > 25$~GeV and $\met > 25$~GeV and the sum of the MC electroweak background predictions, the multijet background prediction, and the MC prediction for signal events.  Systematic uncertainties are not shown.}
\label{fig:MT}
\end{figure}

The background from misidentified multijet events is estimated by fitting the $M_T$  distribution of the $W$ boson candidates with the sum of signal and background shapes. The signal shape is obtained from the same MC simulation as used for the electroweak background. The shape of the multijet background is derived using muon events that fail the isolation criteria under the assumption that the $M_T$ shapes are the same for isolated and non-isolated events. The fit is performed for $50 < M_{T} < 100$~GeV. To determine the systematic uncertainty on the multijet background, we vary the fit range, the $M_{T}$ bin width, and the isolation selection criteria. The largest change in the multijet background is 30\%, which we choose as the systematic uncertainty.  The multijet background is also estimated using several other methods; all give consistent results within similarly large uncertainties.  The multijet background is estimated to be $(3.2 \pm 0.9)$\% of the $W$ boson candidate samples.  The $M_T$ distribution of the selected events is compared with the sum of the background and signal MC events in Fig.~\ref{fig:MT}.

The muon charge asymmetry is also corrected for the muon momentum and \met\ resolutions. This correction is estimated using MC events generated with {\sc resbos+photos}~\cite{resbos,photos} with CTEQ6.6 PDFs~\cite{cteq6.6} and passed through {\sc pythia} for parton showering. The muon momentum and the recoil are then smeared to have the same resolutions as in data~\cite{recoil}. The difference between the asymmetry at the generator level and the asymmetry from the reconstructed MC events (using the same kinematic criteria) is applied to the data to correct for resolution effects.  The shift in the measured asymmetry ranges from nearly zero at $\eta^\mu \approx 0$ to about 12\% of the asymmetry in the largest $|\eta^\mu|$ region analyzed.  A systematic uncertainty due to modeling is included as the difference in the generator-level asymmetries from {\sc resbos+photos} and {\sc powheg}~\cite{powheg} with CT10 PDFs~\cite{CT10}. 

The systematic uncertainty on the muon charge asymmetry is determined from the total uncertainties on the backgrounds, the charge misidentification probability, the relative efficiency for positive and negative muons, the magnet polarity weighting, and the momentum/\met\  resolution correction. A contribution due to varying trigger isolation conditions is also included.  The dominant source of systematic uncertainty is from the momentum/\met\  resolution correction.

\begin{figure}[!ht]
\unitlength1cm
\begin{picture}(8.5,11.7)
\put(-0.45,0.0){\includegraphics[scale=1.0]{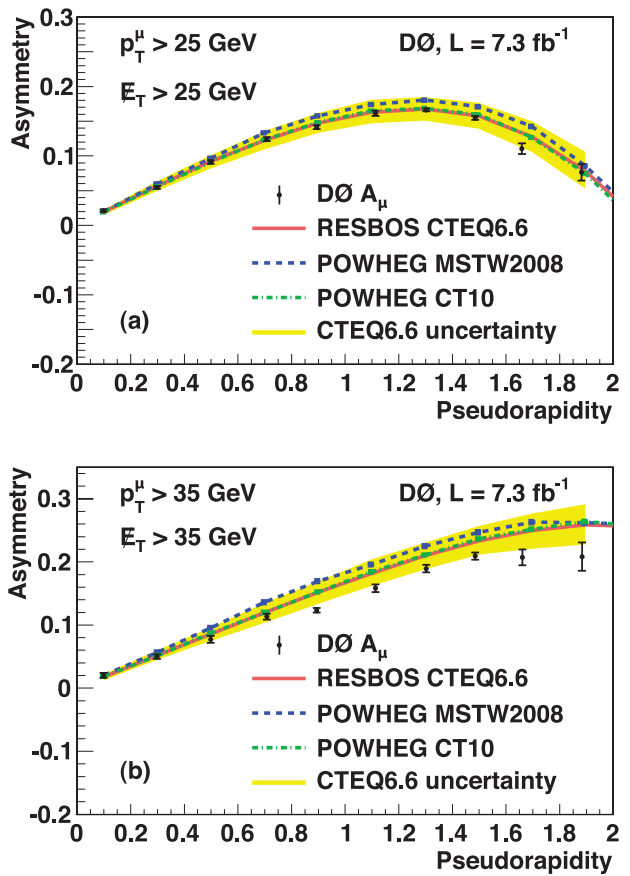}}
\end{picture}
\caption{[color online]. The muon charge asymmetry vs.\ muon pseudorapidity for (a) ($p^{\mu}_{T} > 25$~GeV and $\met > 25$~GeV) and (b) ($p^{\mu}_{T} > 35$~GeV and $\met > 35$~GeV). The black points show the muon charge asymmetry measured with 7.3 fb$^{-1}$ of integrated luminosity.  The error bars represent the total uncertainties.  The solid line and the band are the central value and uncertainty band of the {\sc resbos+photos} with CTEQ6.6 prediction. The predictions from {\sc powheg} with the MSTW2008 and CT10 PDF sets are also shown. }
\label{fig:asymmetry}
\end{figure}

The muon charge asymmetry is expected to be invariant under CP transformation, and our asymmetry results for $\eta^\mu < 0$ are consistent with those for $\eta^\mu > 0$.  Therefore, we fold the data such that $-A_{\mu}(-\eta^\mu) = A_{\mu}(\eta^\mu)$ (CP-folding) to decrease the statistical uncertainty.  The data are CP-folded at the level of the numbers of positive and negative muon events, and all backgrounds, corrections, and uncertainties are remeasured.  Results from Run~IIa and Run~IIb are also found to be consistent and, after CP-folding, combined using the BLUE method~\cite{BLUE}.   Figure~\ref{fig:asymmetry} shows the measured muon charge asymmetry with 7.3~fb$^{-1}$ of integrated luminosity for the two kinematic regions and theory predictions with the CTEQ6.6, CT10, and MSTW2008~\cite{mstw2008} PDF sets. The theory prediction with the CTEQ6.6 PDFs is generated by {\sc resbos+photos}, and the predictions with the CT10 and MSTW2008 PDFs are generated by {\sc powheg}. Both generators are next-to-leading order perturbative QCD calculations interfaced with {\sc pythia} for parton showering. The theory curves are determined by imposing the ($p^{\mu}_{T}$,\met) selection criteria at the generator level. The uncertainty is derived from the CTEQ6.6 uncertainty sets~\cite{cteq-uncerts}.

\begin{table*}[t!]
\caption{\label{Tab:Asym} Muon charge asymmetry for data and predictions from {\sc resbos+photos} using the CTEQ6.6 PDFs. The measurement is shown with statistical uncertainties followed by systematic uncertainties. The uncertainties for the predictions are only from the PDFs.  All asymmetry values are multiplied by~100.}
\begin{ruledtabular}
\begin{tabular}{ccrrrr}
 & & \multicolumn{2}{c}{$p^{\mu}_{T} > 25$~GeV} & \multicolumn{2}{c}{$p^{\mu}_{T} > 35$~GeV}  \\
 & & \multicolumn{2}{c}{$\met > 25$~GeV} & \multicolumn{2}{c}{$\met > 35$~GeV}  \\
$\eta^\mu$ range~~ & $\langle |\eta^\mu| \rangle$ & \multicolumn{1}{c}{$A_{\mu}$} & \multicolumn{1}{c}{Prediction} & \multicolumn{1}{c}{$A_{\mu}$} & \multicolumn{1}{c}{Prediction} \\
\hline
\noalign{\vskip 1mm}
$0.0-0.2$ & $0.10$ & $2.13  \pm 0.17 \pm 0.11$ & $1.97^{+0.28}_{-0.48}$  & $2.03  \pm 0.27 \pm 0.14$ & $1.77^{+0.46}_{-0.53}$  \\
$0.2-0.4$ & $0.30$ & $5.46  \pm 0.18 \pm 0.13$ & $5.68^{+0.71}_{-0.67}$  & $5.01  \pm 0.29 \pm 0.21$ & $5.23^{+0.79}_{-0.74}$  \\
$0.4-0.6$ & $0.50$ & $9.11  \pm 0.18 \pm 0.16$ & $9.24^{+0.86}_{-1.02}$  & $7.71  \pm 0.28 \pm 0.42$ & $8.58^{+1.02}_{-1.11}$  \\
$0.6-0.8$ & $0.71$ & $12.41 \pm 0.18 \pm 0.19$ & $12.23^{+1.33}_{-1.26}$ & $11.34 \pm 0.29 \pm 0.41$ & $11.96^{+1.57}_{-1.58}$ \\
$0.8-1.0$ & $0.89$ & $14.15 \pm 0.19 \pm 0.17$ & $14.76^{+1.42}_{-1.43}$ & $12.32 \pm 0.29 \pm 0.28$ & $15.20^{+1.75}_{-1.85}$ \\
$1.0-1.2$ & $1.11$ & $16.13 \pm 0.16 \pm 0.27$ & $16.29^{+1.81}_{-1.61}$ & $15.84 \pm 0.26 \pm 0.69$ & $18.18^{+2.19}_{-2.00}$ \\
$1.2-1.4$ & $1.30$ & $16.62 \pm 0.14 \pm 0.21$ & $16.76^{+1.71}_{-1.66}$ & $18.94 \pm 0.21 \pm 0.53$ & $21.02^{+2.04}_{-2.20}$ \\
$1.4-1.6$ & $1.49$ & $15.47 \pm 0.16 \pm 0.21$ & $15.78^{+1.90}_{-1.84}$ & $20.92 \pm 0.25 \pm 0.49$ & $23.30^{+2.37}_{-2.17}$ \\
$1.6-1.8$ & $1.66$ & $11.06 \pm 0.70 \pm 0.33$ & $12.75^{+2.26}_{-2.20}$ & $20.71 \pm 1.02 \pm 0.81$ & $24.99^{+2.68}_{-2.90}$ \\
$1.8-2.0$ & $1.88$ & $7.64  \pm 1.07 \pm 0.42$ & $7.83^{+2.75}_{-2.56}$  & $20.83 \pm 1.48 \pm 1.48$ & $25.85^{+3.27}_{-3.11}$ \\
\end{tabular}
\end{ruledtabular}
\end{table*}

At lower lepton $p_{T}$, the lepton charge asymmetry is strongly influenced by the {\sl V--A} decay of the $W$ boson. At large lepton $p_{T}$, the lepton charge asymmetry is closer to the $W$ boson production asymmetry, leading to the different shapes of Figs.~\ref{fig:asymmetry}(a) and \ref{fig:asymmetry}(b).   The data at $p_T^\mu > 35$~GeV, $\met > 35$~GeV, and larger values of $\eta^\mu$ favor an increased ${d(x)}/{u(x)}$ ratio at higher values of $x$ than is predicted, as did the earlier D0 $W\to e\nu$ asymmetry measurement \cite{David}.   The measured values and the {\sc resbos+photos} CTEQ6.6 predictions for both kinematic regions are summarized in Table~\ref{Tab:Asym}.  Contributions of the individual systematic uncertainties are shown in Table~\ref{Tab:sysuncerts-both}.

\begin{table*}[ht!]
\caption{\label{Tab:sysuncerts-both} Contributions from individual sources of systematic uncertainty for the ($p^{\mu}_{T} > 25$, $\met > 25$) [($p^{\mu}_{T} > 35$, $\met > 35$)]~GeV kinematic region.  All uncertainty values are multiplied by 100.}
\begin{ruledtabular}
\begin{tabular}{cccccccc}
 & EW & MJ & Charge & Relative charge & Magnet polarity & Momentum/$\met$ & Trigger \\
 $\eta^\mu$ range & bkg & bkg & mis-id & efficiency & weighting & resolution & isolation \\
\hline
\noalign{\vskip 1mm}
$0.0-0.2~$ & 0.007 [0.004] & 0.018 [0.010] & 0.001 [0.002] & 0.012 [0.012] & 0.006 [0.010] & 0.107 [0.132] & 0.05 [0.04] \\
$0.2-0.4~$ & 0.005 [0.008] & 0.036 [0.034] & 0.006 [0.007] & 0.008 [0.028] & 0.005 [0.008] & 0.129 [0.168] & 0.13 [0.11] \\
$0.4-0.6~$ & 0.029 [0.009] & 0.046 [0.044] & 0.007 [0.010] & 0.013 [0.055] & 0.004 [0.005] & 0.151 [0.402] & 0.06 [0.09] \\
$0.6-0.8~$ & 0.049 [0.039] & 0.065 [0.062] & 0.012 [0.018] & 0.039 [0.084] & 0.003 [0.013] & 0.165 [0.314] & 0.11 [0.23] \\
$0.8-1.0~$ & 0.047 [0.033] & 0.089 [0.059] & 0.012 [0.014] & 0.046 [0.118] & 0.004 [0.010] & 0.134 [0.237] & 0.09 [0.04] \\
$1.0-1.2~$ & 0.051 [0.045] & 0.078 [0.079] & 0.014 [0.017] & 0.053 [0.093] & 0.002 [0.007] & 0.251 [0.614] & 0.22 [0.29] \\
$1.2-1.4~$ & 0.057 [0.074] & 0.058 [0.092] & 0.006 [0.012] & 0.042 [0.103] & 0.002 [0.005] & 0.187 [0.410] & 0.17 [0.29] \\
$1.4-1.6~$ & 0.055 [0.077] & 0.048 [0.101] & 0.013 [0.018] & 0.073 [0.146] & 0.005 [0.008] & 0.183 [0.402] & 0.17 [0.21] \\
$1.6-1.8~$ & 0.030 [0.067] & 0.005 [0.089] & 0.047 [0.133] & 0.082 [0.203] & 0.031 [0.044] & 0.312 [0.534] & 0.20 [0.54] \\
$1.8-2.0~$ & 0.037 [0.085] & 0.009 [0.078] & 0.048 [0.167] & 0.149 [0.418] & 0.049 [0.041] & 0.385 [1.408] & 0.04 [0.04] \\
\end{tabular}
\end{ruledtabular}
\end{table*}

In conclusion, we have measured the muon charge asymmetry from $p\overline{p} \rightarrow W\rightarrow\mu\nu + X$ using 7.3~fb$^{-1}$ of integrated luminosity collected with the D0 detector at $\sqrt{s}=1.96$~TeV. The measured asymmetry is compared with theory predictions generated by {\sc resbos+photos} with the CTEQ6.6 PDF set and by {\sc powheg} with the CT10 and MSTW2008 PDF sets.  The total experimental uncertainties are smaller than the PDF uncertainties in most $\eta^\mu$ regions, so our asymmetry measurement provides additional constraints on the PDFs.  This measurement is a significant improvement on the previous D0 result in this channel and provides the most precise measurement of the $W$ boson lepton asymmetry from the Tevatron for lepton pseudorapidities $|\eta^\ell| \lesssim 1.8$.

%
We thank the staffs at Fermilab and collaborating institutions,
and acknowledge support from the
DOE and NSF (USA);
CEA and CNRS/IN2P3 (France);
MON, NRC KI and RFBR (Russia);
CNPq, FAPERJ, FAPESP and FUNDUNESP (Brazil);
DAE and DST (India);
Colciencias (Colombia);
CONACyT (Mexico);
NRF (Korea);
FOM (The Netherlands);
STFC and the Royal Society (United Kingdom);
MSMT and GACR (Czech Republic);
BMBF and DFG (Germany);
SFI (Ireland);
The Swedish Research Council (Sweden);
and
CAS and CNSF (China).

\end{document}